\documentclass[%
 aip,
 amsmath,amssymb,
 reprint,%
]{revtex4-1}

\usepackage{graphicx}
\usepackage{dcolumn}
\usepackage{bm}

\usepackage[utf8]{inputenc}
\usepackage[T1]{fontenc}
\usepackage{mathptmx}
\usepackage{etoolbox}
\usepackage{xcolor}

\makeatletter
\def\@email#1#2{%
 \endgroup
 \patchcmd{\titleblock@produce}
  {\frontmatter@RRAPformat}
  {\frontmatter@RRAPformat{\produce@RRAP{*#1\href{mailto:#2}{#2}}}\frontmatter@RRAPformat}
  {}{}
}%
\makeatother
\begin{document}

\preprint{AIP/123-QED}

\title{Dynamics of cascades in spatial interdependent networks}
\author{Bnaya Gross}
 \affiliation{Department of Physics, Bar-Ilan University, 52900 Ramat-Gan, Israel.}
 \email{bnaya.gross@gmail.com}
 \author{Ivan Bonamassa}
\affiliation{Department of Network and Data Science, CEU, Quellenstrasse 51, 1100 Vienna, Austria}
\email{ivan.bms.2011@gmail.com}
\author{Shlomo Havlin}%
  \affiliation{Department of Physics, Bar-Ilan University, 52900 Ramat-Gan, Israel.}

\date{\today}

\begin{abstract}
The dynamics of cascading failures in spatial interdependent networks significantly depend on the interaction range of dependency couplings between layers. In particular, for increasing range of dependency couplings, different types of phase transition accompanied by various cascade kinetics can be observed including mixed-order transition characterized by critical branching phenomena, first-order transition with nucleation cascades, and continuous second-order transition with weak cascades. We also describe the dynamics of cascades at the mutual mixed-order resistive transition in interdependent superconductors and show its similarity to that of percolation of interdependent abstract networks. Finally, we layout our perspectives for the experimental observation of these phenomena, their phase diagrams and the underlying kinetics, in the context of physical interdependent networks. Our studies of interdependent networks shed light on the possible mechanisms of three known types of phase transitions, second order, first order, and mixed order as well as predicting a novel fourth type where a microscopic intervention will yield a macroscopic phase transition. 
\\
\\
\textbf{In honor of Prof. Juergen Kurths’ 70th birthday}
\end{abstract}

\maketitle

\begin{quotation}
The theory of interdependent networks has been developed to describe dependency relations between infrastructures and to understand their resilience, the propagation of cascading failures, and the conditions leading to the abrupt collapse of such systems. Interdependent networks are characterize by self-amplifying cascading processes fueled by the positive feedback induced by dependency couplings with critical dynamics that generally depend on the network topology. The theory has been motivated by improving the understanding of interdependent infrastructures such as power grids and their communication systems. However, the theory could not be proved in real-world systems since infrastructures are not possible to control. The recent experimental realization of interdependent networks as thermally coupled disordered superconductors, hereafter called physical interdependent networks (PINs) for brevity, has allowed for the first time the manifestation under a controlled environment of self-amplifying cascade dynamics analogue to those observed in interdependent percolation on abstract structures, raising new perspectives in the study of coupled macroscopic systems. Here we lay out the analogies between the various types of cascade dynamics reported in both abstract and physical interdependent networks and provide our vision for future studies.
\end{quotation}

\section{Introduction}
A common feature of biological \cite{junker2011analysis,alm2003biological}, technological \cite{barabasi-science1999,albert1999diameter}, ecological \cite{montoya2006ecological,bascompte2010structure}, and social \cite{borgatti2018analyzing} systems is the ability to represent many of them as networks. The ability to abstract a complex system by nodes and edges representing their interactions, without losing its important features is one of the significant advantages of the complex networks paradigm and the reason for its interdisciplinary applications. About a decade ago, researchers realized that networks in various areas are not isolated but rather interact and depend on each other and that a theory for such system of systems was missing. This understanding has led to the development of the paradigm of \textit{interdependent networks} \cite{buldyrev2010catastrophic, vespignani2010fragility,parshani2010interdependent,gao2012interdependentnetworks} followed by a large variety of models like multiplex networks~\cite{nicosia2013growing, battiston2014structural}, network of networks~\cite{d2014networks, bianconi2014multiple,radicchi2017redundant} and unifying frameworks for the structure and function of multilayer networks~\cite{de2013mathematical,de2015structural,de2016physics}. 
\par
\begin{figure}[b]
	\centering
{\includegraphics[width = \linewidth]{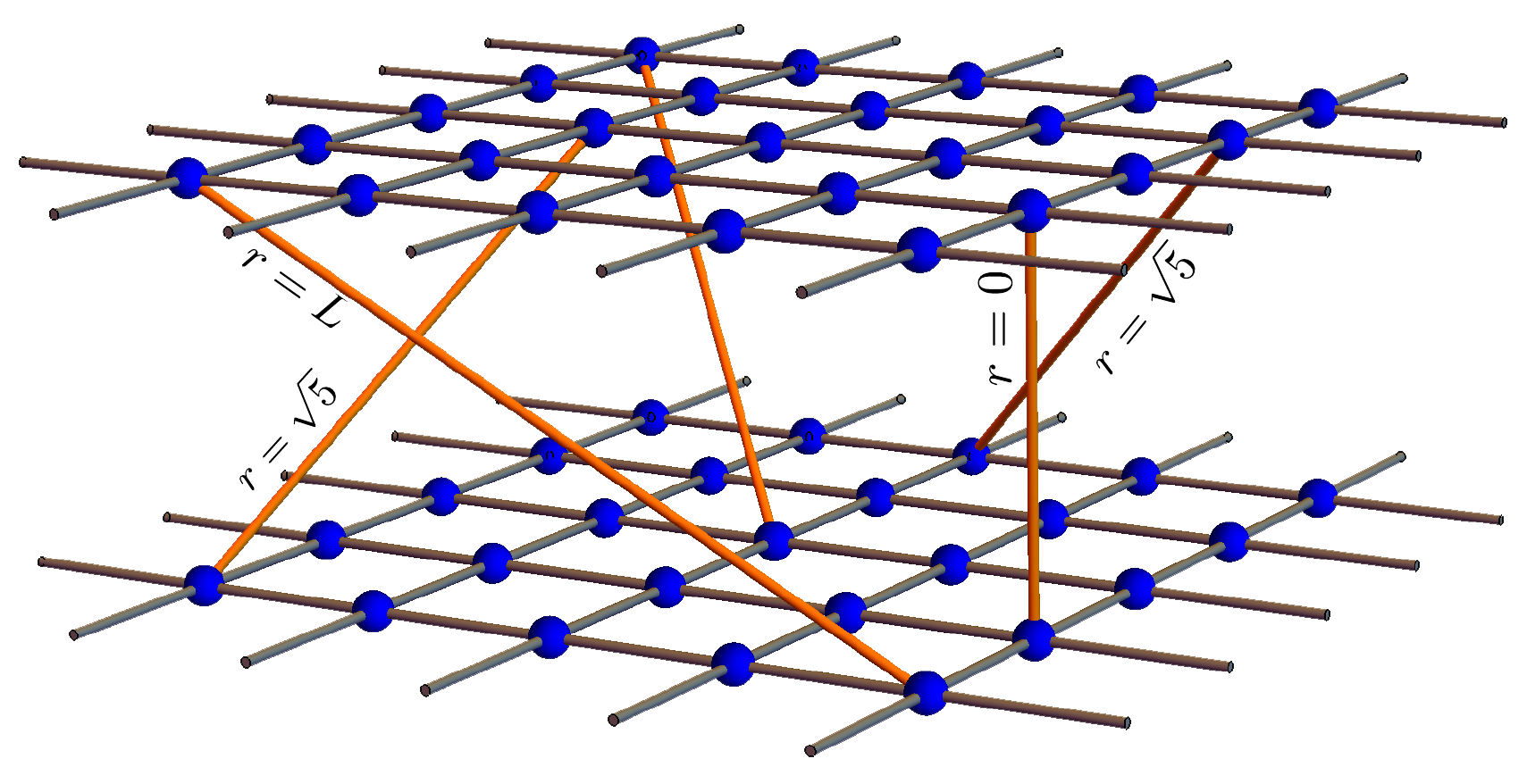}}\vspace*{-0.25cm}
	\caption{\textbf{Abstract interdependent networks.} Illustration emphasizing the presence of two qualitatively different types of links: connectivity links (gray lines) within each network and dependency links (orange lines) between them. Here the network structure is taken to be a square lattice of linear size $L$ in $d=2$ but it could be a lattice in any dimension or a random graph. Dependency links can be assigned randomly within a radius $r$.}
	\label{fig:illustration}	
\end{figure}

Interdependent networks, in particular, have the distinctive feature of modeling systems endowed with two types of couplings: \textit{connectivity} links within layers and \textit{dependency} links between them, as illustrated in Fig.~\ref{fig:illustration}. The role of each type of links is different: while connectivity links are used to describe the structural connectivity of the network for its specific function, dependency links are used to describe functional dependence between components in different networks so that, e.g.\ failures can propagate between them \cite{buldyrev2010catastrophic}. As a result, the interplay of connectivity and dependency links offers a simple mechanisms describing the positive feedback triggering the catastrophic phenomena reported in power outages~\cite{rosato-criticalinf2008, yang2017small} or cascading tipping points in critical infrastructures~\cite{rinaldi-ieee2001, ferrari2023vulnerability} and ecosystems~\cite{scheffer2003catastrophic, rocha2018cascading, scheffer2020critical,pocock-science2012}. These cascades are self-amplifying processes \cite{motter2017unfolding} initiated by microscopic perturbations that, when close to a critical point, can lead the global shifts of the system's state. Works focusing on interdependent percolation in spatially embedded networks~\cite{wei-prl2012, bashan2013extreme, berezin-scireports2015, gao2011robustness} have revealed the vulnerability of these structures to external microscopic localized failures, disclosing a variety of kinetic regimes. In this paper, we review some key properties of these cascading processes in the presence of constraints on the range of dependency links. We then highlight the novel phenomena emerging when studying cascading failures in physical interdependent networks, offering future perspectives of experimental validation. The theory of percolation of interdependent networks sheds light on the mechanisms of three types of known phase transitions, first order, mixed order, and second order. While the second order transition occurs when both interactions (connectivity and dependency couplings) are short range, mixed order transition  occurs when one or both interactions  are long-range of the order of the system size \cite{parshani2010interdependent,buldyrev2010catastrophic}. The first-order abrupt transition occurs   due to random nucleation when one coupling is short range and the other is of length shorter than the system size \cite{wei-prl2012,berezin-scireports2015,danziger-epl2016}. Surprisingly, the theory of percolation phase transition of interdependent network predicts a novel phase regime of macroscopic phase transition that occurs due to microscopic intervention \cite{berezin-scireports2015} (see also Fig.\ref{fig:pc_r}\textbf{c} below).

\begin{figure*}
	\centering
{\includegraphics[width=\textwidth]{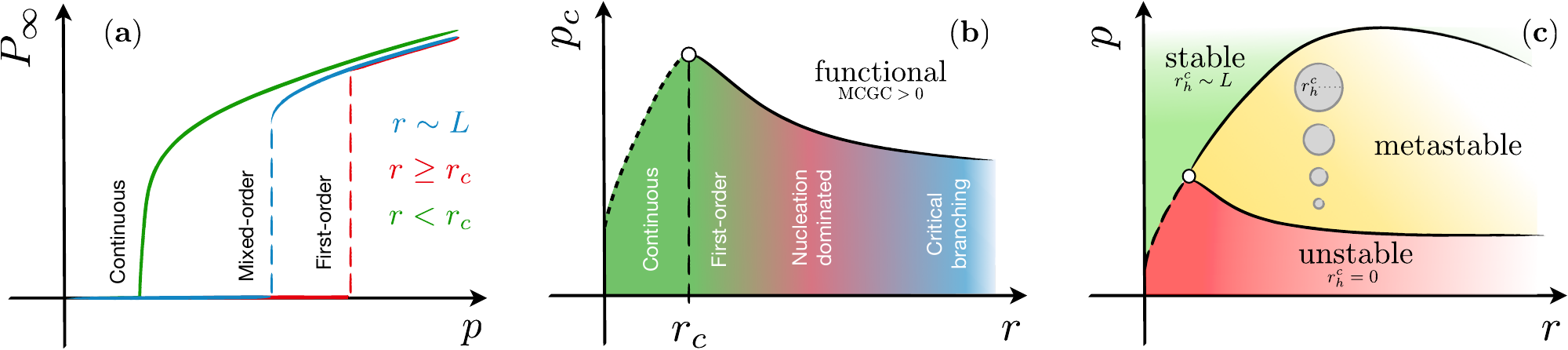}}
	\caption{\textbf{Transitions and interaction range. (a) } Three types of transitions are observed depending on the interaction range \cite{wei-prl2012}: continuous for short-range ($r < r_c$), first-order due to nucleation for intermediate-range ($r_c \leq r \ll L$), and mixed-order for long-range ($r \sim L$).  \textbf{(b)} This behavior is reflected by the value of the critical point, $p_c$, as the interaction range varies. For short-range $r < r_c$ the transition is continuous and $p_c$ increases linearly with $r$. For an intermediate range above $r_c$, abrupt first-order transition due to spontaneous nucleation is observed, and $p_c$ decreases. For long-range interaction, $r \sim L$, mixed-order transition is observed and $p_c$ is converging to its long-range interaction limit \cite{wei-prl2012}. (\textbf{c}) Three phases are observed, both numerically and theoretically \cite{berezin-scireports2015}: an unstable phase (red) where the system spontaneously breaks; a stable phase where the system can not be destroyed by microscopic intervention (green), and a metastable phase  where the system can be destroyed via microscopic intervention. In the metastable phase microscopic localized attacks anywhere in the system, of size above a critical radius, $r_c^h$, dismantle the system. At the critical line between unstable and stable phases (dashed line), the transition is continuous. In contrast, at the critical line between the unstable and metastable phases (full line), the transition is abrupt first-order, and the cascading dynamics is governed by nucleation kinetics. The empty circle indicates a tricritical point. In the metastable regime (yellow) the size of the  localized microscopic critical attack, $r_h^c$, changes with $p$ and $r$.}
	\label{fig:pc_r}	
\end{figure*}

\section{Vulnerability of interdependent networks}
A practical approach to characterize the vulnerability and the propagation of failures in interdependent structures is percolation theory \cite{staufferaharony,fan2020universal,bunde1991fractals}. Let us start by briefly describing the basic case of percolation in a single isolated network. In the percolation process, a fraction of $1-p$ of nodes are randomly removed from the network and the relative size of the largest (giant) connected component (GCC), $P_{\infty}$, is measured. The GCC describes the connectivity of the network and its existence is regarded as a meaningful proxy for the functionality of the network. A percolation transition is commonly observed below a critical threshold, $p_c$, where the network itself breaks apart into small clusters and the relative size of its giant connected component, $P_{\infty}$, vanishes. For a classical percolation process, cascades are typically absent and the transition is usually continuous, as depicted in Fig.~\ref{fig:pc_r}\textbf{a}. 
\par
In marked contrast to percolation of isolated networks, interdependent percolation on coupled networks exhibits different and richer phenomena. In this framework, one usually starts from the random removal of a fraction $1-p$ of nodes from one of the networks, after which its remaining GCC is measured. Notice that, since the GCC is the functional part of the network, small clusters concurrently fail. At this stage, the dependency links transmit the failures of these nodes and of small disconnected clusters to the other network(s). In their turn, these failures disconnect some clusters from the GCC of the other network, propagating new failures through the dependency links back to the first network. As this process iterates back and forth, \textit{cascade of failures} propagate between the layers until either the entire system is dismantled or a stable mutual giant connected component (MGCC)---a subset of the giant connected components of both layers composed by the functional nodes in the GCC of both networks---remains. When the external damage is sufficiently large, these cascades result in abrupt mixed or first-order percolation transitions, depending on the range of interactions, as displayed in Fig.~\ref{fig:pc_r}\textbf{a}. 

The surprising feature of the change in the transition's order relies on the underlying kinetics of failures generated by the dependency couplings. In fact, it was shown that the dynamics of cascades is characterized by different critical features which strongly depend on the range of interactions \cite{wei-prl2012}. In what follows we will focus on the effects that a limited range, $r$, of the dependency couplings has on the kinetics of cascading failures in the simple model of two interdependent lattices depicted in Fig.~\ref{fig:illustration}.

\section{The role of the dependency interaction range}

To study how the range of dependency links affects the observed phase transition, a spatial interdependent network model was developed \cite{wei-prl2012}. In this model (shown in Fig.~\ref{fig:illustration}), two 2D square lattices are interdependent on each other and the dependency links are constrained to be below a specific geometric range $r$.  For $r=0$, percolation of interdependent networks is identical to that of a single network, with $p_c=0.593$. This is because failures in one network yield identical failures in the second network and there will be no feedback of cascades.  In the limiting case of very short-range dependencies---say, of a few lattice units (see Fig.~\ref{fig:illustration})---cascades propagate only locally and the percolation transition remains continuous \cite{wei-prl2012}. As the dependency range increases, the critical threshold, $p_c$, also increases without though influencing the character of the transition in the system (see Fig.~\ref{fig:pc_r}\textbf{b}). Wei Li {\em et al}.\ showed \cite{wei-prl2012} the existence of a critical value $r_c$, whose value is close to the value of the correlation length of a single system, above which avalanches propagate in a nucleating fashion. In this case, above $r_c$, the transition occurs when a small droplet of damage is spontaneously created at the critical threshold, $p_c$, and the dependency links amplify it by spreading it radially until the entire system collapses abruptly. In this case, in sharp contrast to the second-order phase transition, no critical scaling is observed in the relative size of the giant component.


As $r$ further increases above $r_c$, the critical threshold, $p_c$,  decreases, reaching eventually the asymptotic regime of $p_c$ for $r\sim L$, where $L$ is the linear size of the lattice. In this case, dependency links become long-range, critical droplets become more and more {\em ramified}~\cite{heermann1983nucleation} and the transition crosses over from nucleation-dominated to mixed-order, exhibiting scaling exponents near the critical threshold $p_c$ and fractal fluctuations phenomena~\cite{gross2022fractal}. In this limit of long-range dependencies, cascades are typically characterized by a {\em critical branching} process with branching ratio $\eta\sim1$, a microscopic property of the kinetics which reflects itself in a long-lived metastable plateau stage observed in the evolution of the MGCC~\cite{dong-pre2014}.



\begin{figure*}
	\centering
{\includegraphics[width = 0.75\linewidth]{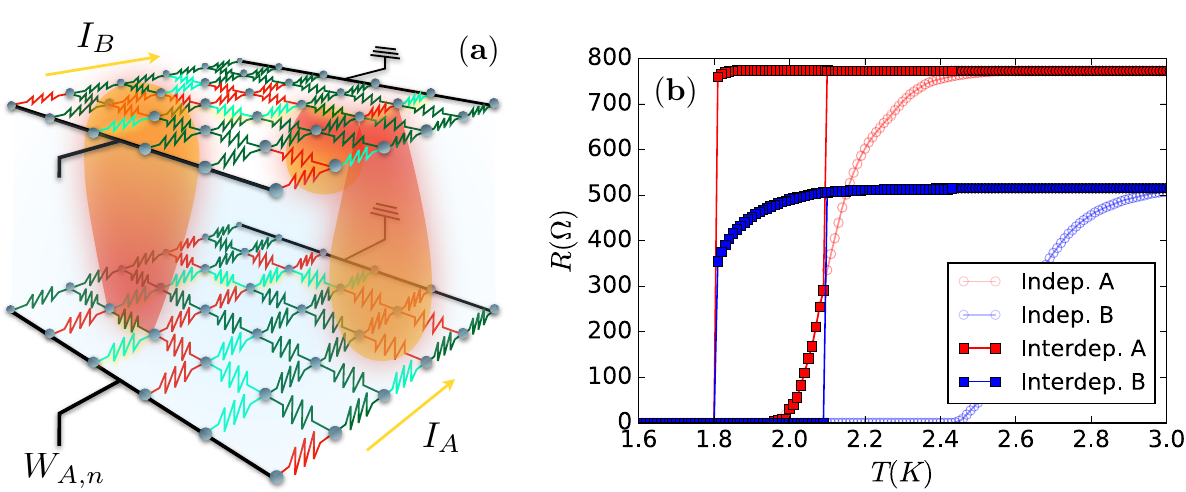}}
	\caption{\textbf{Interdependent superconducting networks.} (\textbf{a}) Model representation of the experimental setup \cite{bonamassa2023superconductors} via a network of two thermally coupled resistively-shunted Josephson junctions (RSJJs) organized in a lattice geometry. Notice that the lattices are disordered since each junction is endowed by its own critical current, $I_c$, and critical temperature, $T_c$---both, randomly distributed---at which a superconducting-normal phase transition occurs. (\textbf{b}) In isolation, each network experiences a second-order (continuous) transition at a distinct critical point. Once the networks are thermally coupled, the transition becomes abrupt with a joint critical point for both networks, and hysteresis is observed. }
	\label{fig:supercondcutors_transitions}	
\end{figure*}

\section{Localized attack}

Fig.~\ref{fig:pc_r}\textbf{a} exhibits three types of phase transitions that appear in interdependent networks under random failures. The second order transition occurs when both interactions (connectivity and dependency couplings) are short-range, mixed order transition  occurs when one or both are long-range of the order of the system size \cite{parshani2010interdependent,buldyrev2010catastrophic,vaknin-njp2017}. The first order transition occurs due to random nucleation when one coupling is short range and the other is of length larger than $r_c$ but shorter than the system linear size \cite{wei-prl2012,berezin-scireports2015,danziger-epl2016}. The theory of interdependent networks also predicts a novel fourth macroscopic phase transition which is triggered via a microscopic intervention. This fourth type of structural transition also depends on the range of the interdependent interactions. This transition can be regarded as a \textit{nucleation-induced transition} since it results from the spontaneous propagation of a microscopic droplet of removed modes whose size (Fig.~\ref{fig:pc_r}\textbf{c}), remarkably, encompasses only a vanishing fraction of the system size \cite{berezin-scireports2015}. This form of percolation process was presented in the literature under the term of ``localized attack'' since it is typically initiated by removing nodes within a circle of radius $r_h$ anywhere in the system. In simulations, for simplicity and without loss of generality, localized attacks are performed in the center of one of the coupled networks. At a given value of $p$ above the spontaneous nucleation critical line, when the network is connected, a critical radius size $r_h^c$ exists where for a localized attack of $r_h > r_h^c$ the damaged hole will propagate and destroy the system while for $r_h < r_h^c$ it will remain local (Fig.~\ref{fig:pc_r}\textbf{c}). It is important to note that $r_h^c$ does not depend on the system size and therefore can be regarded as a microscopic intervention that yields a macroscopic phase transition \cite{berezin-scireports2015}. The regime in which a microscopic intervention yields a macroscopic phase transition (marked in yellow) is called the metastable regime. This is because the system is not really fully stable since a microscopic intervention, anywhere in the system, yields the collapse of the system. This process enables to probe the stability of the MGCC of interdependent lattices, unveiling an upper bound in the phase diagram of the model (see the yellow area, Fig.~\ref{fig:pc_r}\textbf{c}) where the coupled are structurally metastable. Notice that, a critical exponent describing the scaling of the critical radius of the droplet with the average degree of the underlying networks has been reported in the metastable regime \cite{vaknin-njp2017}.


\section{Cascade kinetics in interdependent superconducting networks}
On the one hand, interdependent percolation on coupled networks has helped to understand some of the key mechanisms underlying cascading failures in real-world systems, however, the ability to test and further develop its predictions in laboratory-controlled experiments has been missing. To fill this fundamental gap, we have recently conducted an experiment performed on thermally-coupled disordered superconductors~\cite{bonamassa2023superconductors}, where heat dissipation physically realizes the dependency coupling. In this experiment, two superconducting networks (illustrated in Fig.~\ref{fig:supercondcutors_transitions}\textbf{a}) are placed on top of each other with an electrically isolated material in between which has good thermal conductivity. When the networks are measured separately, each layer experiences a continuous superconductor-normal (SN) transition, as shown in Fig.~\ref{fig:supercondcutors_transitions}\textbf{b}. However, once the layers are coupled, thermal interactions set in between the layers via dissipating hotspots which trigger electro-thermal runaway effects that cause the layers to lock in their critical temperature, eventually leading to mutually abrupt superconducting-normal phase transitions.
\par

\begin{figure}
	\centering
{\includegraphics[width = \linewidth]{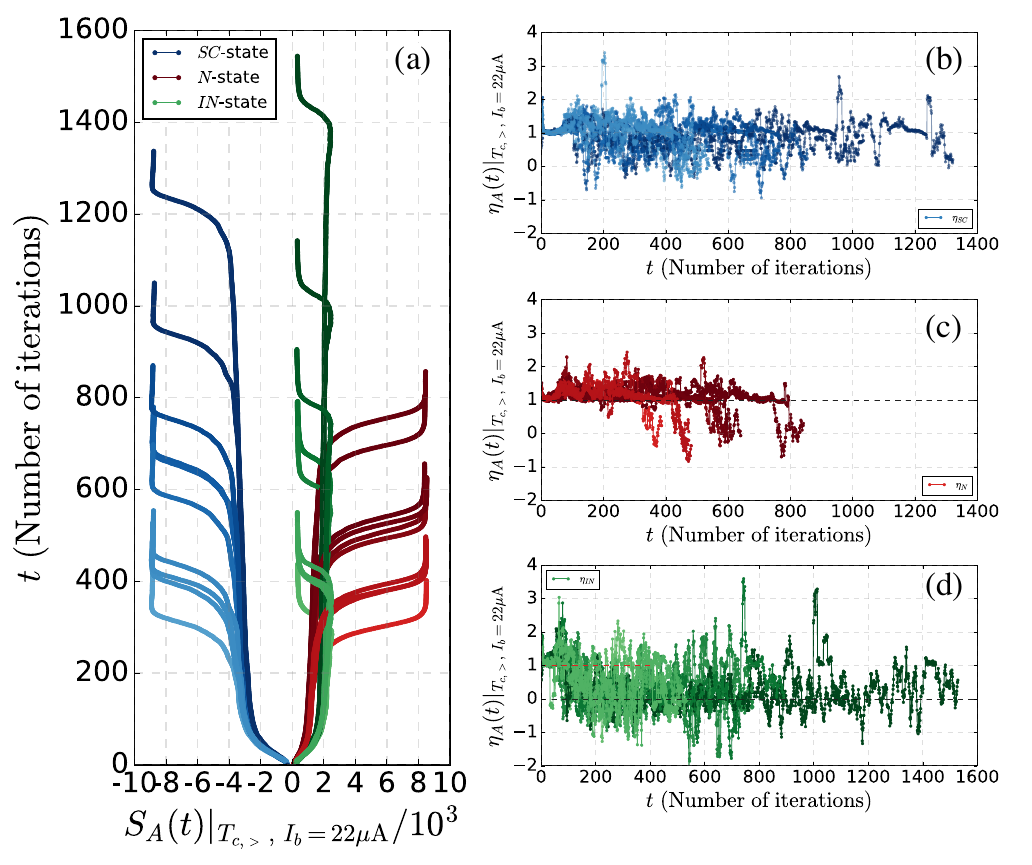}}
	\caption{\textbf{Plateau lifetime of cascades in interdependent superconductors.} (\textbf{a}) Evolution of the size of N-state cascades (red symbols), intermediate (IN)-state (green symbols) and SC-state (blue symbols) at the N-to-SC transition threshold. (\textbf{b}-\textbf{c}-\textbf{d}) Evolution of the branching factor, $\eta(t)=S(t+1)/S(t)$ for (\textbf{b}) SC-avalanches, (\textbf{c}) N-avalanches and (\textbf{d}) IN-avalanches.  Note that while the N/SC-avalanches appear to be critical during their evolution, the IN-branching factor is here clearly neutral ($\eta_{IN}\simeq0)$. See Ref.~\cite{bonamassa2023superconductors} for details.}
	\label{fig:supercondcutors_plateau}	
\end{figure}
In order to characterize this phenomenon and its connection with interdependent percolation, we have developed a model of thermally coupled $2D$ resistively-shunted Josephson junctions (RSJJs), where local dissipation is modeled via a local, Joule heating effect (see illustration in Fig.~\ref{fig:supercondcutors_transitions}\textbf{a}). In particular, we have modeled the state of a given lattice bond, $(i,j)$, via a Josephson I-V characteristics featuring one of three possible states: superconductor (SC), intermediate (I) and normal (N). These states are defined by the junction’s critical current $I_{ij}^c$ and its normal-state resistance $R_{ij}^n$, whose values depend on the local temperature, $T_{ij}$. We describe the latter via a local de Gennes relation~\cite{de1976relation}
\begin{equation}
    I_{ij}^c (T_{ij}) = I_{ij}^c(0)\big(1-T_{ij}T_{ij}^c\big)^2 ,
    \label{eq:deGennes}
\end{equation}
where $I_{ij}^c(0)$ is the zero-temperature critical current of the junction and $T_{ij}^c$ is its activation temperature, whose values are extrapolated from the experimental data. To measure the global resistances of the networks as a function of temperature and of the bias current, we solved numerically the Kirchhoff equations $\mathbf{G} \cdot \mathbf{W} = \mathbf{I}_b$ for each layer, where $\mathbf{G}$ is the conductance matrix, $\mathbf{W}$ is the potential vector and $\mathbf{I}_b$ is the current vector. To model the thermal coupling between the two superconducting networks, we have calculated at each iteration in the numerical solution of the Kirchhoff equations, the power dissipated by Joule heating of single junctions, i.e.\ $P_{ij, t}=R_{ij}^2 I_{ij, t}$, where $I_{ij,t}$ is the current passing through the junction $(i,j)$ at the $t$-th numerical iteration. An \textit{effective} local temperature can then be obtained by thermal circuit arguments so to take into account the mutual overheating effect between the networks. In particular, given the much larger thermal conductance between the layers than within layers, one can write the local expression~\cite{bonamassa2023superconductors}
\begin{equation}\label{eq:1}
T_{ij,t}^\mu =T+\frac{\tau_p}{\tau_e}\gamma^{-1}P_{ij,t-1}^{\mu},
\end{equation}
\noindent
where $\gamma\,[\mathrm{WK}^{-1}]$ is the thermal conductance of the coupling medium and $\mu'\neq\mu$, with $\mu,\mu'=\mathrm{A},\mathrm{B}$. In Eq.~\eqref{eq:1}, the ratio $\tau_p/\tau_e$ between the two relevant time scales ($\tau_p$ for phonons and $\tau_e$ for electrons) characterize the heat rate transferred through the coupling medium and the one emitted by Joule dissipation, have values that generally depend on the geometry of the sample as well as on the physical properties of the superconducting materials. 

Given the local overheating effect induced by Eq.~\eqref{eq:1}, we solved iteratively the coupled Kirchhoff equations characterizing the thermally coupled RSJJs. At zero temperature all bonds are superconductors and no dissipation is present. As the temperature increases, the critical current of bonds decreases according to Eq.~\eqref{eq:deGennes} and some of them switch their state from superconducting (SC) to dissipating (IN or N). At sufficiently large currents, these bonds overheat the other layer, increasing the ``vulnerability'' of the latter ones to switch as well to the normal state. At sufficiently large currents, a critical temperature $T_c$ of the heat bath is eventually reached, at which the local overheating effect between the networks couples with the electrical runaway within layers, causing local perturbations to be propagated at large scales. When this electrothermal feedback process is ignited, more and more bonds switch to the normal state and a mutually abrupt resistive transition is observed in both layers (see Fig.~\ref{fig:supercondcutors_transitions}\textbf{b}).


\begin{figure}
	\centering
{\includegraphics[width = \linewidth]{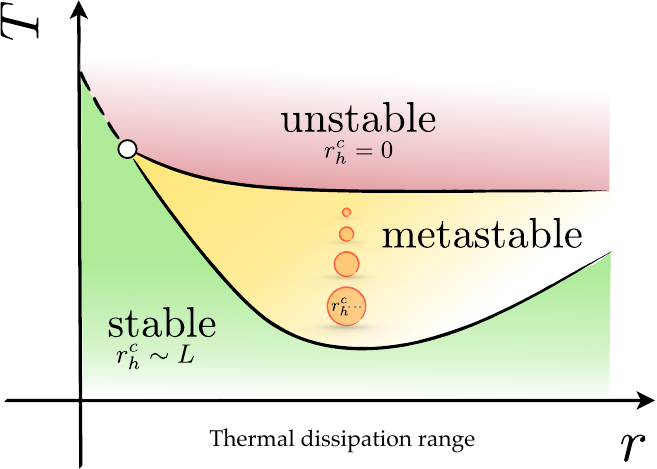}}
	\caption{\textbf{Our perspective.} We expect localized microscopic heating and cooling of PINs to realize and further extend interdependent percolation with richer phase diagrams and cascading kinetics, whose perspective we briefly sketched. The mutual normal and superconductor phases have been observed theoretically and experimentally only in the limits of single networks ($r \to 0$) and long-ranged coupled networks ($r \to \infty$). However, the intermediate dependence range is yet to be explored and a mutual metastable phase is expected to appear therein, together with novel intertwined kinetics. This perspective is based on the results obtained for percolation of abstract interdependent networks shown in Fig.~\ref{fig:pc_r}c.}
	\label{fig:vison}	
\end{figure}

Interestingly, the critical kinetics underlying the two abrupt (i.e.\ mutual SC-to-N and N-to-SC phase) transitions are accompanied by different relaxation processes. At the mutual SC-to-N transition, the \textit{overheating cascade} process physically realizes the kinetics of cascading failures of interdependent percolation~\cite{zhou2014simultaneous}, further manifested by the classical long-lived \textit{plateau} stage  whose lifetime $\tau\propto(T-T_{c,>})^{-\zeta}$ with $\zeta\simeq0.65$ diverges at $T_{c,>}$. In the cooling direction, on the other hand, while the evolution from the mutual N-phase to the mutual SC-phase exhibits an analogous plateau regime (Fig.~\ref{fig:supercondcutors_plateau}\textbf{a}) its characteristic lifetime diverges at the N-to-SC threshold, $T_{c,<}$, as $\tau\propto(T_{c,<}-T)^{-\zeta}$ now with exponent $\zeta\simeq0.5$ (for details, see Ref.~\cite{bonamassa2023superconductors}. For percolation of interdependent networks $\zeta=0.5$, see Ref.~\cite{zhou2014simultaneous}). 
Microscopically, the different critical exponents of the plateau lifetimes can be adopted as proxies for the underlying cascading kientics~\cite{binder1987theory} \!, indicating that the SC-nuclei grow faster then N-nuclei. 
During the heating plateau, this can be explained in terms of the pinning of the interfaces between SC-clusters and N-nuclei which halts the branching of the latter, while the smaller exponent of the cooling plateau hints at the sudden merging of thermally-suppressed SC-clusters. The critical nature of these dynamics is reflected in the evolution of the cascading trees generated by state-switching junctions (Fig.~\ref{fig:supercondcutors_plateau}). At the transition temperatures, in fact, the avalanche size $S(t)$, i.e.\ the number of junctions cascading to the SC/N-state at time $t$, develops a long-lived plateau (Fig.~\ref{fig:supercondcutors_plateau}\textbf{a}) during which its relative growth is a zero fraction of the system's size and a critical branching factor $\eta_c \sim 1$ is typically observed (Fig.~\ref{fig:supercondcutors_plateau}\textbf{b,c}). 

\section{Future perspectives}
Interdependent networks~\cite{buldyrev2010catastrophic} feature rich and unique dynamics of cascades that governs their macroscopic phase transition, resulting in dramatic changes in the type of transitions from mixed-order to nucleation-dominated or continuous. The spatial range of dependency/connectivity couplings, in particular, plays a key role in this respect, as vividly embodied by the so-called interdependent $r$-model~\cite{wei-prl2012, bashan2013extreme,berezin-scireports2015} and the so-called multiplex $\zeta$-model \cite{vaknin-njp2017,danziger-epl2016} discussed above. The recent realization of PINs as interdependent superconducting networks~\cite{bonamassa2023superconductors} offers the opportunity of controlling and validating in experiments a large body of theoretical and numerical results gathered in the context of interdependent spatial networks~\cite{vespignani2010fragility,danziger-epl2016, vaknin2017spreading,wei-prl2012, bashan2013extreme,berezin-scireports2015}. Furthermore, the appearance of four different types of phase transitions in a single model improves our understanding of the mechanisms of phase transitions in general. Nonetheless, the expected fourth type of induced nucleation transition is novel and has yet to be observed in PINs. In our vision, a phase diagram of \textit{localized heating} (Fig.~\ref{fig:vison}) should be studied both theoretically and experimentally completing the picture of phase transitions in PINs.

\section{Acknowledgments} 
We thank the Israel Science Foundation, the Binational Israel-China Science Foundation Grant No.\ 3132/19, NSF-BSF Grant No.\ 2019740, the EU H2020 project RISE (Project No. 821115), the PAZY Foundation, and the EU H2020 DIT4TRAM. B.G. acknowledges the support of the Mordecai and Monique Katz Graduate Fellowship Program.

\section{Authors Decleration}
\noindent
\textbf{Conflict of Interest} \\
\\
\quad The authors have no conflicts to disclose.\\
\\
\textbf{Data Availability} \\
\\
\quad The data that support the findings of this study are available
from the corresponding author upon reasonable request..\\
\\
\par
\noindent
\textbf{Author Contributions} \\
\\
\textbf{Bnaya Gross}: Conceptualization (equal); Investigation; Validation (equal); Visualization (equal); Writing – original draft (equal). \textbf{Ivan Bonamassa}: Conceptualization (equal); Investigation; Visualization (equal); Validation (equal); Writing – original draft (equal). \textbf{Shlomo Havlin}: Conceptualization (equal); Supervision (equal); Project administration (equal); Validation
(equal); Writing – review \& editing (equal).

\bibliographystyle{unsrt}
\bibliography{mybib}
\end{document}